\documentclass{emulateapj}
\usepackage{apjfonts}

\slugcomment{August 6, 2007; revised June 30, 2008; accepted August 11, 2008}

\shorttitle{GeV EMISSION FROM GAMMA-RAY BURSTS}
\shortauthors{ANDO, NAKAR, \& SARI}

\begin{document}

\title{G\lowercase{e}V Emission from Prompt and Afterglow Phases of
Gamma-Ray Bursts}
\author{Shin'ichiro Ando, Ehud Nakar, and Re'em Sari}
\affil{California Institute of Technology, Mail Code 130-33, Pasadena,
CA 91125;\\
ando@tapir.caltech.edu, udini@tapir.caltech.edu, sari@tapir.caltech.edu}

\begin{abstract}
We investigate the GeV emission from gamma-ray bursts (GRBs), using
the results from the Energetic Gamma Ray Experimental Telescope
(EGRET), and in view of the {\it Gamma-ray Large Area
Space Telescope (GLAST)}. Assuming that the conventional prompt and
afterglow photons originate from synchrotron radiation, we compare
an accompanying inverse-Compton component with EGRET measurements
and upper limits on GeV fluence, taking Klein-Nishina
feedback into account. We find that EGRET constraints
are consistent with the theoretical framework of the synchrotron
self-Compton model for both prompt and afterglow phases, and
discuss constraints on microphysical parameters in both phases.
Based on the inverse-Compton model and using EGRET results, we
predict that {\it GLAST} would detect GRBs with GeV
photons at a rate $\gtrsim 20$ yr$^{-1}$ from each of the prompt and
afterglow phases. This rate applies to the high-energy tail of
the prompt synchrotron emission and to the inverse-Compton component
of the afterglow. Theory predicts that in a large fraction of the
cases where synchrotron GeV prompt emission would be detected by
{\it GLAST}, inverse-Compton photons should be detected as well at high
energies ($\gtrsim 10$ GeV). Therefore {\it GLAST} will enable a more
precise test of the high-energy emission mechanism. Finally, we show
that the contribution of GRBs to the flux of the
extragalactic gamma-ray background measured with EGRET is at least
0.01\% and likely around 0.1\%.
\end{abstract}

\keywords{gamma-rays: bursts --- radiation mechanisms: non-thermal}

\section{Introduction}
\label{sec:Introduction}

Cosmological gamma-ray bursts (GRBs) have released a tremendous
amount of energy in the past and present Universe. Their emission
covers very wide range of frequencies: a highly variable prompt
phase radiates $\sim$100 keV gamma rays, while a subsequent
afterglow radiates radio to X-ray photons. It is likely that the
bulk of these photons are emitted by gyration of relativistic
electrons in magnetic fields---e.g., synchrotron radiation.
The relativistic electrons are accelerated in
either internal dissipation (for prompt emission) or external shocks
(for afterglows).
For reviews, see, \citet{Piran2005,Meszaros2006,Nakar2007}.

GeV photons were detected as well from several GRBs by the Energetic
Gamma Ray Experimental Telescope (EGRET) on board the {\it Compton
Gamma Ray Observatory (CGRO)}
\citep{Schneid1992,Sommer1994,Hurley1994,Schneid1995,Gonzalez2003}.
The data are still not sufficient for us to firmly infer emission
mechanisms of these GeV gamma rays, but the most promising mechanism
is synchrotron self-Compton (SSC) scattering
\citep*[e.g.,][]{Meszaros1994,Waxman1997,Wei1998,Chiang1999,
Panaitescu2000,Zhang2001,Sari2001,Guetta2003}. This is because the
relevant emission parameters such as the energy fraction of the GRB jets
going to electrons ($\epsilon_e$) and magnetic fields ($\epsilon_B$)
are relatively well measured from the afterglow spectra as well as
light curves; the typical values are $\epsilon_e = 0.1$ and
$\epsilon_B = 0.01$ \citep[e.g.,][]{Panaitescu2001,Yost2003}. In the
prompt emission, $\epsilon_e$ is similar or even higher, as evident
from the high efficiency of this phase, while $\epsilon_B$ is not
well constrained.  Thus, there should be a significant
inverse-Compton (IC) component accompanying the synchrotron
radiation in both the afterglow and prompt emission. The
luminosities of the synchrotron and IC are expected to be comparable
as IC-to-synchrotron luminosity ratio is roughly given by
$(\epsilon_e / \epsilon_B)^{1/2}$, according to theory
\citep[e.g.,][]{Sari2001}.

In this paper, we explore the GeV gamma-ray emission of GRBs in
the context of SSC mechanism.\footnote{Our analysis and
conclusions are applicable also if the MeV and/or radio-X-ray
afterglow emission mechanism is not synchrotron but another type of
emission from relativistic electrons that gyrate in a magnetic
field, such as jitter radiation \citep{Medvedev00}.}  Besides the
several GRBs detected by EGRET, there are many others for which
upper bounds on the fluence were obtained \citep{Gonzalez2005}.
These $\sim$100 GRBs should also be compared with the predictions of
SSC model, because the fluence upper limits in the EGRET energy band
are comparable to the fluence of prompt emission collected by
Burst And Transient Source Experiment (BATSE) instrument onboard
{\it CGRO}. As the experimental bound is already strong, while
theoretical models of SSC process predict a large fluence for the
EGRET energy range, we derive meaningful constraints from EGRET data
analysis on the physics of the high-energy emission mechanisms of
GRBs. This approach is different from (and therefore complementary
with) that in previous studies
\citep*[e.g.,][and references therein]{Dermer00, Asano2007, Ioka2007,
Gou2007, Fan2008, Murase2008, Panaitescu2008},
where the prediction of
gamma-ray flux relies only on theoretical models and sub-GeV
observations. We instead use EGRET data in order to infer the GeV
emission and constrain the theoretical models.

We use our results to predict the expected number of GRBs that
would be detected by the {\it Gamma-ray Large Area Space
Telescope (GLAST)}. The {\it GLAST} satellite is equipped with the
Large Area Telescope (LAT), which is an upgraded version of EGRET.
Since revealing the
high-energy emission mechanisms of GRBs are one of the important
objectives of {\it GLAST}, our prediction should give a useful
guideline. Finally, we apply our results to estimate the
contribution of GRBs to the diffuse extragalactic gamma-ray
background (EGB), which was also measured by EGRET \citep[][see,
however, \citealt*{Keshet2004b} for a subtle issue of Galactic
foreground subtraction]{Sreekumar1998,Strong2004}.

This paper is organized as follows. In \S~\ref{sec:IC}, we
summarize the predictions of SSC model for the prompt
(\S~\ref{sub:prompt}) and afterglow
(\S~\ref{sub:afterglow}) phases. Section~\ref{sec:Constraint on
high-energy emission with EGRET} is devoted for analysis of the GRB
fluence data by EGRET, from which distributions of fluence in the
GeV band are derived. We then use these distributions to argue
prospects for GRB detection with {\it GLAST} in
\S~\ref{sec:GLAST}, and implications for EGB from GRB emissions
in \S~\ref{sec:EGB}. In \S~\ref{sec:conclusions}, we give
a summary of the present paper.

\section{Inverse-Compton model of high-energy emission}
\label{sec:IC}

If the prompt and/or afterglow emission is due to synchrotron
radiation from relativistic electrons (with Lorentz factor
$\gamma_e$), then there must be an accompanying IC component from
the same electrons scattering off the synchrotron photons. The
spectral shape of the IC emission is almost the same as the
synchrotron radiation (shifted by $\gamma_e^2$), and is expected to
fall around the GeV range during both the prompt and afterglow
phases. For $\epsilon_e > \epsilon_B$, and assuming that there is no
``Klein-Nishina suppression'' and that the emitting electrons are
fast cooling, the IC fluence is related to the synchrotron fluence
simply through $F_{\rm IC} \approx (\epsilon_e / \epsilon_B)^{1/2}
F_{\rm syn}$. Thus, assuming that the microphysics do not vary much
from burst to burst, it is natural to assume proportionality between
the synchrotron MeV fluence (observed by BATSE) and the GeV
synchrotron plus IC fluence (observed by EGRET and in the future by
{\it GLAST}):
\begin{equation}
 F_{\rm GeV} = (\eta_{\rm syn} + \eta_{\rm IC}) F_{\rm MeV},
 \label{eq:EGRET fluence}
\end{equation}
where $\eta_{\rm syn}$ and $\eta_{\rm IC}$ are coefficients for the
proportionality due to synchrotron and IC processes.
Note that the synchrotron fluence in the GeV range can be extrapolated
relatively easily, if we assume that the spectrum extends up to such
high energies.
Thus, we here focus on theoretical evaluation of the IC component.
At first approximation, the coefficient $\eta_{\rm IC}$ is roughly
$(\epsilon_e / \epsilon_B)^{1/2}$ from considerations above,
and thus we define
\begin{equation}
 \eta_{\rm IC} = \left(\frac{\epsilon_e}{\epsilon_B}\right)^{1/2}
  \xi_{\rm KN} \xi_w \frac{F_{\rm syn}}{F_{\rm MeV}},
  \label{eq:eta_IC}
\end{equation}
where for the prompt emission $F_{\rm syn} \approx F_{\rm MeV}$
while for the afterglow $F_{\rm syn}$ is the afterglow fluence
within the radio to X-ray energy bands. Correction factors $\xi_{\rm
KN}$ and $\xi_w$ represent the effect of Klein-Nishina suppression and
detector energy window, respectively, which are given below.

We define typical frequencies for both synchrotron ($\nu_{\rm syn}$)
and IC ($\nu_{\rm IC}$) as the frequencies where most of the
energies are radiated  in case that the Klein-Nishina
cross section does not play an important role; i.e., where $\nu
f_\nu$ for each component is peaked in this case. From relativistic
kinematics, these two typical frequencies are related through
\begin{equation}
 \nu_{\rm IC} \approx \gamma_m^2 \nu_{\rm syn},
  \label{eq:frequency}
\end{equation}
where $\gamma_m$ is a characteristic Lorentz factor of the electrons
that dominate the synchrotron power \citep{Rybicki1979}; this is true
in the fast cooling regime, which is the case in the most of our
discussions \citep{Sari2001}. The Klein-Nishina effect is relevant if a
photon energy in the electron rest frame exceeds the electron rest mass
energy, and this condition is formulated as
\begin{equation}
 h\nu_{\rm KN} =
  \Gamma_b \gamma_m m_e c^2,
 \label{eq:Klein-Nishina}
\end{equation}
where $\Gamma_b$ is the bulk Lorentz factor of the ejecta,
which is on the order of 100 in the prompt phase of GRBs and their
early afterglows. Upscattering synchrotron
photons to energies above $h \nu_{\rm KN}$ is highly suppressed,
which results in IC cutoff at $\nu_{\rm KN}$.

Besides producing a spectral cutoff, the Klein-Nishina effect also
modifies the way electrons cool, which is relevant for the GeV
emission and is also included in $\xi_{\rm KN}$. Electrons with
energies above Klein-Nishina threshold (for a given seed-photon
energy) can lose their energies only through synchrotron radiation,
while the lower-energy ones can cool through both processes. Such an
effect has been studied in the case where the seed photons for IC
scattering are provided by an external sources \cite[e.g.,][and
references therein]{Moderski2005a,Moderski2005b}. However, in the
case of SSC mechanism, since the seed photons are emitted from
synchrotron process due to the same electron population, we should
properly take into account feedback. Giving full details on this is
beyond the scope of the present paper,
but some results are summarized briefly in Appendix~\ref{app:KN}
(see also \citealt{Derishev03}). Here we only show the approximate
analytic form of $\xi_{\rm KN}$:
\begin{equation}
 \xi_{\rm KN} \approx
  \left\{
   \begin{array}{ccc}
    1 & \mbox{for} & \gamma_m \le \gamma_{\rm KN}, \\
    \left(\frac{\gamma_m}{\gamma_{\rm KN}}\right)^{-1/2}
     & \mbox{for} & \gamma_m > \gamma_{\rm KN}, \\
   \end{array}
  \right.
  \label{eq:eta KN}
\end{equation}
where $\gamma_{\rm KN}$ is the Lorentz factor of electrons for
which photons at $\nu \gtrsim \nu_{\rm syn}$ are in the
Klein-Nishina regime. The energy of an observed photon with
frequency $\nu$ as measured in the rest frame of an electron with
Lorentz factor $\gamma$ is $\approx \gamma h\nu/\Gamma_b$ where the
$1/\Gamma_b$ factor converts the photon energy from the observer
frame to the plasma rest frame and the $\gamma$ factor converts it
to the electron rest frame. Since such a photon is in the
Klein-Nishina regime of an electron with Lorentz factor $\gamma$
once its energy in the electron rest frame is larger than $m_e c^2$
we obtain:
\begin{equation}
 \gamma_{\rm KN} = \frac{\Gamma_b m_e c^2}{h\nu_{\rm syn}}.
  \label{eq:gamma KN}
\end{equation}
This Klein-Nishina feedback effect modifies the spectrum shape of both
synchrotron and IC emissions (in addition to the Klein-Nishina
cutoff for IC).
We note that equation~(\ref{eq:eta KN}) provides a solution that agrees
within a factor of $\sim$2 with the one obtained by numerically
solving equation~(\ref{eq:Y}).
This precision is sufficiently good for our purpose, especially because
it is well within the uncertainty ranges of other parameters.

By $\xi_w$, we take into account the fraction of the IC fluence that
falls into the GeV detector energy bands. EGRET window is between $h
\nu_{w,l}=30$ MeV and $h \nu_{w,u}=30$ GeV while {\it GLAST}-LAT
window is between $h \nu_{w,l}=20$ MeV and $h \nu_{w,u}=300$ GeV. We
here assume that the frequency where most of the IC energy is
released, $\nu_{\rm IC, peak} \equiv \min[\nu_{\rm IC}, \nu_{\rm
KN}]$, is always larger than lower limit of the frequency band,
$\nu_{w,l}$, as expected for both EGRET and {\it GLAST}, and thus
consider the cases in which $\nu_{\rm IC,peak}$ is within or above
the detector frequency band. In
the former case where $\nu_{w,l} < \nu_{\rm IC, peak} < \nu_{w,u}$,
we have $\xi_{w} \approx 1$. On the other hand, if $\nu_{\rm IC,
peak} > \nu_{w,u}$, then most of the energy comes from the upper
frequency limit $\nu_{w,u}$, and we have $\xi_{w} \approx (\nu_{w,u}
/ \nu_{\rm IC, peak})^{2 - \alpha_1}$, where $\alpha_1$ is the photon
spectral index below peak frequency.  Thus we may approximate
$\xi_{w}$ as
\begin{eqnarray}
 \xi_{w} & \approx &
  \left(1+\frac{\min[\nu_{\rm IC},\nu_{\rm KN}]}
  {\nu_{w,u}}\right)^{\alpha_1 - 2}
  \nonumber\\ &=&
  \left\{
   \begin{array}{ccc}
    \left(1+\frac{\gamma_{m}^2 \nu_{\rm syn}}{\nu_{w,u}}
     \right)^{\alpha_1 - 2} & \mbox{for} &
     \gamma_{m} \le \gamma_{\rm KN},\\
    \left(1+\frac{\Gamma_b\gamma_{m}m_e c^2}
      {h\nu_{w,u}}\right)^{\alpha_1 - 2} & \mbox{for} &
     \gamma_{m} > \gamma_{\rm KN},\\
   \end{array}
  \right.
 \label{eq:eta_w 1}
\end{eqnarray}
where $\nu_{\rm IC} / \nu_{\rm KN} = \gamma_m / \gamma_{\rm KN}$ as one
can easily show.

The discussion above assumes that the density of the
synchrotron photon field is proportional to the instantaneous
synchrotron emissivity. In the case of relativistically expanding
radiation front, this assumption is valid when the duration over
which the emissivity vary significantly, $\delta t$, is comparable
to the time that passed since the expanding shell was ejected, $t_0$.
In this case the ratio between the synchrotron emissivity and the
synchrotron photon field density is in a steady state. When $\delta
t \ll t_0$ the synchrotron photon field density may be significantly
lower than in the steady state case \citep*{Granot2007}, thereby
suppressing the IC component. The exact suppression factor depends
on the detailed spatial and temporal history of the emissivity.
Theoretically, in the afterglow phase we expect $\delta t
\sim t_0$. Also in the prompt emission phase, internal shock models
generally predict $\delta t \sim t_0$ \cite[][and references
therein]{Piran99}. Thus, in the internal-external shock model
corrections to the IC component due to this effect are expected to
be on the order of unity. Therefore, in the present paper, we assume
that such an effect can be neglected and that the synchrotron photon
field is proportional to the instantaneous synchrotron emissivity.
One should keep in mind, however, that $\delta t \ll t_0$ is a viable
possibility
\citep[see, e.g.,][for a more detailed study in such cases]{Peer2004,
Peer2005},
especially in the highly variable prompt
phase. In principle, detailed GLAST observations of an IC emission may
be able to constrain $\delta t /t_0$ during the prompt phase.

In addition, towards the higher end of the EGRET or {\it GLAST}
energy band, photons may start to be subject to absorption due to
pair creation in the source or during propagation
\citep*[e.g.,][]{Baring1997, Lithwick2001, Razzaque2004, Ando2004,
Casanova07, Murase2007}. Although such a mechanism might be relevant
for the IC yields (especially in the prompt phase) depending on some
parameters that are not well constrained yet, we assume that it is not
the case in the present paper. {\it GLAST} will hopefully provide
information that enables better handle on this issue.

\subsection{Prompt phase}
\label{sub:prompt}

\begin{figure}
\begin{center}
\includegraphics[width=8.6cm]{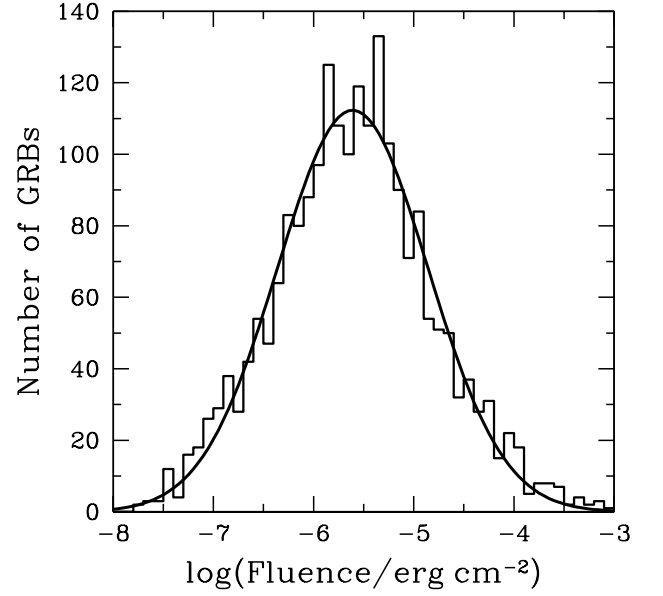}
\caption{The fluence distribution in the prompt phase of BATSE
 GRBs. The best-fit log-normal function is also shown, where the peak is
 at $\overline F_{\rm BATSE} = 2.5 \times 10^{-6}$ erg cm$^{-2}$
 and the standard deviation is $\sigma_{\log F} = 0.75$.}
\label{fig:fluence}
\end{center}
\end{figure}

BATSE (as well as {\it Swift} satellite) detected so far a large
number of GRBs in prompt phase with gamma rays in the energy band
of 20 keV--1 MeV. The spectrum is well described by a smoothly
broken power law with a typical lower-energy index of $\alpha_1
\approx 1$ and higher-energy index of $\alpha_2 \approx 2.3$; the
spectral break typically occurs around $h\nu_{\rm syn} \approx 300$
keV, where the energy of the prompt emission $\nu f_\nu$ peaks
\citep{Band1993,Preece2000,Kaneko2006}. As we show in
Figure~\ref{fig:fluence}, the distribution of the fluence integrated
over the BATSE energy band follows log-normal
function.\footnote{http://www.batse.msfc.nasa.gov/batse/grb/} The
peak of this distribution is $\overline F_{\rm BATSE} = 2.5 \times
10^{-6}$ erg cm$^{-2}$, and its standard deviation is $\sigma_{\log
F} = 0.75$. The average of the BATSE fluence is therefore $\langle
F_{\rm BATSE} \rangle = 10^{-5}$ erg cm$^{-2}$.

Therefore, for the prompt emission phase, using $\alpha_1 = 1$,
$\alpha_2 = 2.3$, and $h \nu_{\rm syn} = 300$ keV, we find:
\begin{equation}
 \gamma_{\rm KN} = 170 \Gamma_{b,2}.
  \label{eq:break Lorentz factor}
\end{equation}
where $\Gamma_{b,2}= \Gamma_{b}/10^2$.
In addition, for $\xi_w$, considering {\it GLAST}-LAT energy window (20
MeV--300 GeV) in equation~(\ref{eq:eta_w 1}), we obtain
\begin{equation}
 \xi_{w} =
  \left\{
   \begin{array}{ccc}
     \left[1+\left(\frac{\gamma_{m}}{10^3}\right)^2\right]^{-1}
     & \mbox{for} & \gamma_{m} \le 170\Gamma_{b,2},\\
     \left(1+\frac{\Gamma_{b,2} \gamma_{m}}{5900}\right)^{-1}
     & \mbox{for} & \gamma_{m} > 170 \Gamma_{b,2}.
   \end{array}\right.
    \label{eq:eta_w 2}
\end{equation}
Now assuming that all electrons are accelerated in the shocks, the
typical value for the Lorentz factor of the relativistic electrons are
given as
\begin{equation}
 \gamma_{m} \approx \epsilon_{e}
  \frac{m_{p}}{m_{e}}
  (\Gamma_{\rm rel}-1)
  = 200 \epsilon_{e,-1} (\Gamma_{\rm rel}-1),
  \label{eq:gamma_e}
\end{equation}
where $\Gamma_{\rm rel}$ is the relative Lorentz factor of the colliding
ejecta portions and $\epsilon_{e,-1} = \epsilon_e / 10^{-1}$. In the
internal shock model for the prompt
emission, $\Gamma_{\rm rel} - 1$ is of order unity. If we adopt
$\Gamma_{\rm rel} = 3$ and $\epsilon_{e} = 0.1$, we obtain
$\gamma_{m} \approx 400$. Furthermore, assuming $\Gamma_{b}
= 100$, equation~(\ref{eq:eta_w 2}) gives $\xi_w \approx 0.9$, and
equation~(\ref{eq:eta KN}) with equation~(\ref{eq:break Lorentz factor})
gives $\xi_{\rm KN} \approx 0.7$.
By substituting these values and assuming $\epsilon_B = 0.01$ in
equation~(\ref{eq:eta_IC}), we obtain $\eta_{\rm IC} \approx 1.9$, which
implies that under the most straightforward assumptions a comparable
fluence is expected in both {\it GLAST}-LAT and BATSE windows.
In this case, the Klein-Nishina cutoff energy is in {\it GLAST}-LAT band
as well as in EGRET band ($h \nu_{\rm KN} \lesssim 30$ GeV), and thus we
also obtain another comparable value of $\eta_{\rm IC} \approx 1.2$ in
EGRET case.

Note that in the case of prompt emission, the synchrotron spectrum
is not negligible in EGRET and {\it GLAST}-LAT energy bands. For
canonical parameters ($h \nu_{\rm syn} = 300$ keV, $\alpha_1 = 1$,
and $\alpha_2 = 2.3$), the ratio $\nu f_{\nu,{\rm IC}} / \nu
f_{\nu,{\rm syn}}$ at 100 MeV is about $0.01 (\epsilon_{e} /
\epsilon_B)^{1/2} (\gamma_m / 400)^{-2}$, assuming that the
synchrotron spectrum continues into the GeV window without a break
and IC is not much suppressed by the Klein-Nishina effect.
Therefore, the synchrotron component dominates around the
lower-energy limit where most of the photons (although not most of
the fluence) are observed. In the case of EGRET, since only a
handful of photons were detected in all EGRET events, these are
expected to be dominated by the synchrotron low-energy ($\sim$100
MeV) photons. This indicates that the quantity we can constrain
using the EGRET fluence upper limits is not $\eta_{\rm IC}$ but
$\eta_{\rm syn} = F_{\rm syn}(100 ~\mathrm{MeV}) / F_{\rm MeV}$, the
ratio of synchrotron fluence around 100 MeV and that in the MeV
range. In addition, this picture is indeed consistent with the fact
that the spectral indices of GeV photons for several GRBs measured
with EGRET are $\alpha = 2$--3
\citep[e.g.,][]{Schneid1992,Sommer1994,Hurley1994}.   Note however,
that the {\it energy fluence} in {\it GLAST}-LAT and EGRET bands can
be dominated by a much harder IC component ($\alpha \approx 1$--2)
that peaks above $\sim$1 GeV and may carry up to $\sim$10 times more
energy than the one observed at 100 MeV without being detected.
This is because even when the $\sim$10 GeV fluence is ten times
larger, the small photon number at such high-energies is still small
enough to avoid detection.  Thus, EGRET observations, which are
consistent with measurement of the synchrotron high energy tail, can
only put an upper limit on $\eta_{\rm IC}$.

\subsection{Afterglow phase}
\label{sub:afterglow}

The afterglow is considered to be a synchrotron emission from
electrons accelerated in the external shock, which is caused by the
interaction between the relativistic ejecta and the interstellar
medium. In this model, the synchrotron emission dominates the
spectrum from radio to X-ray. The associated IC emission is expected
to dominate the GeV energy range (i.e., $\eta_{\rm IC} \gg \eta_{\rm
syn}$), since the electron Lorentz factor is much larger than the case
of prompt emission (see eq.~[\ref{eq:gamma_e}], where the relative and
bulk Lorentz factors are the same, $\Gamma_{\rm rel} = \Gamma_{b}$),
compensating the smaller $\nu_{\rm syn}$
(eq.~[\ref{eq:frequency}]). During the first several minutes
(observer time), electrons might be cooling fast ($\alpha_1 = 1.5$)
with $h\nu_{\rm syn} \approx 1$ keV, while $\gamma_{m} \approx
10^4$--10$^5$. This implies that the fraction of the IC energy that
falls in {\it GLAST}-LAT  energy window is close to unity, i.e,
$\xi_{w} \approx 0.2$--0.9 from equation~(\ref{eq:eta_w 1}) (for EGRET
$\eta_{w} \approx 0.08$--0.5) and $\xi_{\rm KN} \approx 0.7$--1 from
equations~(\ref{eq:eta KN})--(\ref{eq:gamma KN}). Since $h\nu_{\rm IC}$ at
early time is close to the upper limit of the energy window the
effective photon index of the IC emission within the detector window
during this time is $\approx$1.5--2.

At later times the electrons are at the slow-cooling regime and
$\nu_{\rm syn}$ is the cooling frequency, while a typical $\gamma_{e}$
is the Lorentz factor of electrons that cooled significantly
\citep[e.g.,][]{Sari2001}. In this regime the SSC peak is very broad
and its location is almost constant with time. For typical
parameters, the Klein-Nishina effect do not play a major role while
the peak of the SSC emission falls within {\it GLAST}-LAT and EGRET
windows. Therefore, at late time $\xi_w \approx 1$ and the
effective photon index within the energy windows of these detectors
is $\approx$2.

One should, however, note that on long time scales the GeV background
becomes important, making it hard to detect the GeV afterglow.
Therefore, the optimal time scale for GeV afterglow search would be
$\sim$100--10$^3$ s \citep{Zhang2001}. The afterglow GeV fluence,
$F_{\rm GeV}$ in equation~(\ref{eq:EGRET fluence}), is that
integrated over a given time scale, while $F_{\rm MeV}$ is collected
over roughly $T_{90}$, during which 90\% of the MeV photons are
counted. The total energy radiated away by the radio to X-ray
afterglow during every decade of time is roughly 0.01--0.1 of the
energy emitted in the prompt phase. Therefore we expect a bright GeV
afterglow which radiate about 
$0.01$--$0.1 (\epsilon_{e}/\epsilon_B)^{1/2} F_{\rm MeV}$ every
decade of time for hours and days after the bursts. In this paper
when considering EGRET observations, we adopt 200 s after $T_{90}$,
when electrons are in the fast cooling regime, as the duration over
which $F_{\rm GeV}$ is integrated.

\section{Constraint on high-energy emission with EGRET}
\label{sec:Constraint on high-energy emission with EGRET}

\citet{Gonzalez2005} analyzed GRBs that were detected by BATSE and
observed by EGRET. Since the field of view of EGRET was much smaller
than that of BATSE and the observation was limited by the life time
of the spark chamber, EGRET covered only about 100 GRBs out of
$\sim$3000 BATSE bursts. But this is still a reasonably large number
to get statistically meaningful result. The analysis of the prompt
burst in EGRET data was performed around the error circles of BATSE
bursts for the first $T_{90}$, and spectral index of $-2.4$ is
assumed within EGRET window (the upper limits are higher by a factor
of $\approx$10 for a spectral index of $-1$). The same analysis was
performed for the afterglow phase, for 200 s after $T_{90}$ (not
including $T_{90}$). \citet{Gonzalez2005} measured the fluence of 6
and 12 GRBs, in prompt and afterglow phases respectively. For all
other GRBs only fluence upper limits were obtained in the range
$10^{-6}$--$10^{-3}$ erg cm$^{-2}$.

\begin{figure}
\begin{center}
\includegraphics[width=8.6cm]{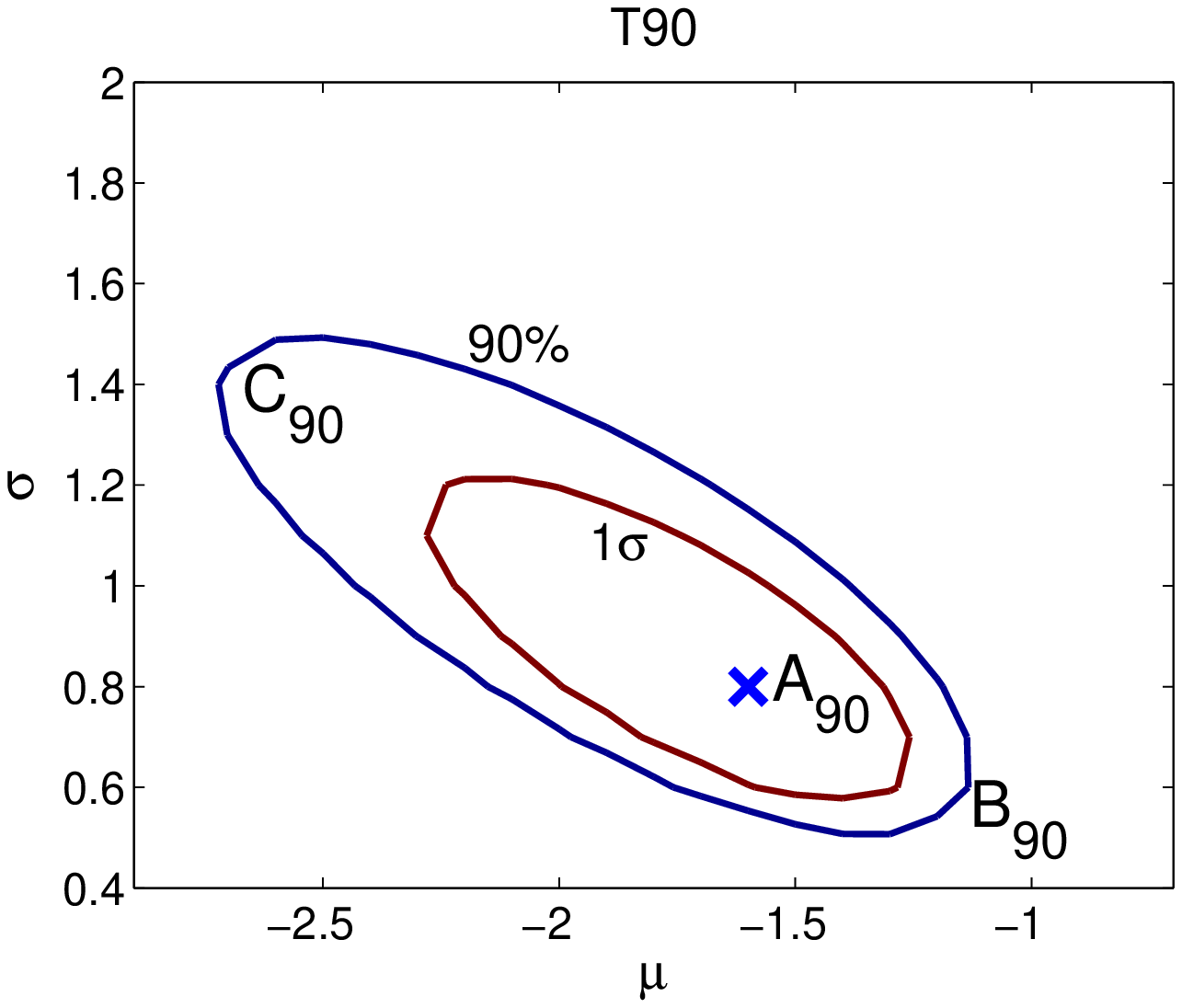}
\includegraphics[width=8.6cm]{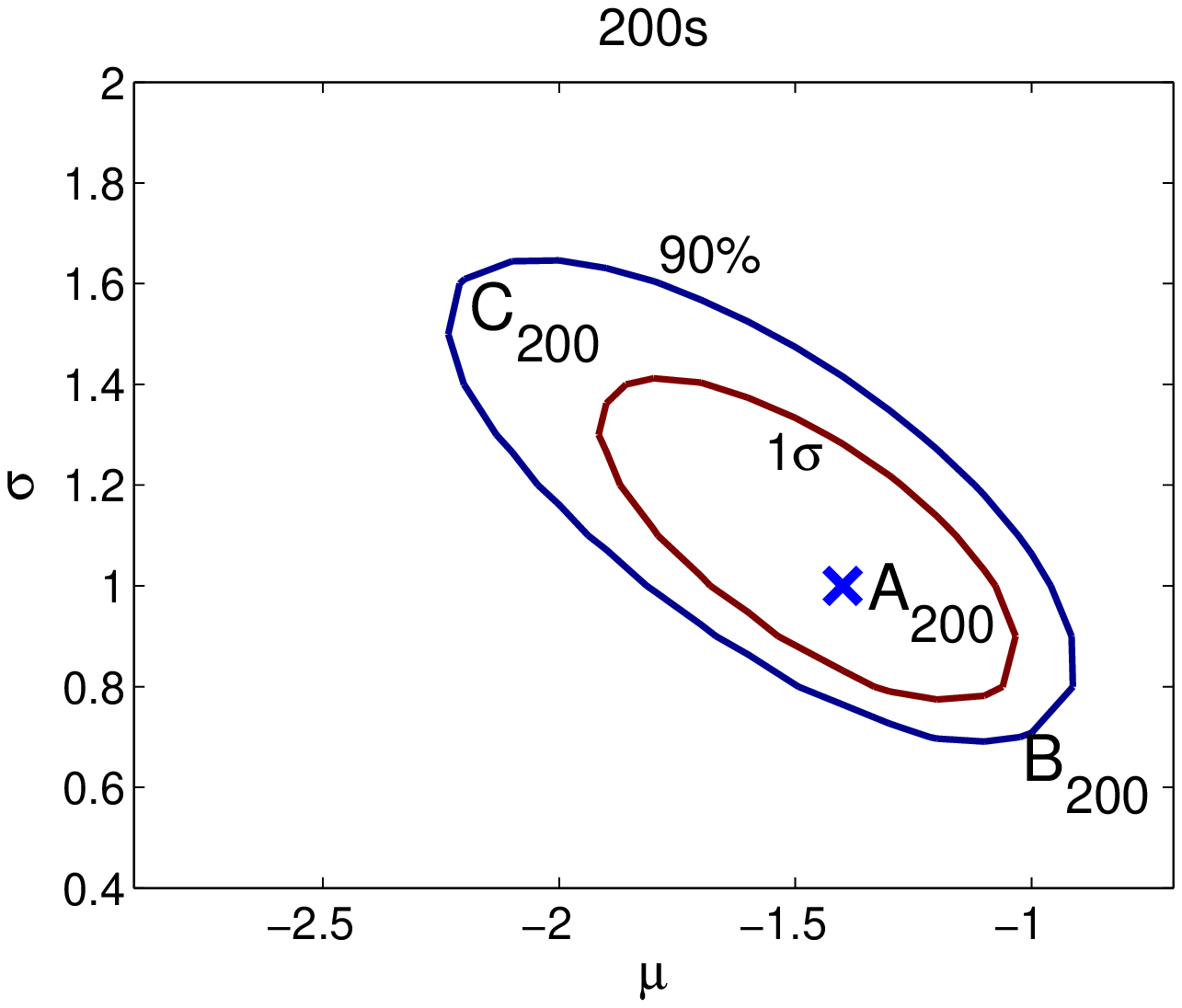}
\caption{Contour plot of allowed region in $\mu$--$\sigma$ space,
 obtained with the analysis of EGRET data assuming a spectral index of
 $-2.4$ during $T_{90}$ ({\it top}) and 200 s after $T_{90}$ ({\it
 bottom}).
 $\mu$ and $\sigma$ are the central value and standard
 deviation, respectively, for the log-normal distribution of the
 fluence ratio $\eta$ (eq.~[\ref{eq:eta distribution}]).
 The best fit
 points (A) are marked as crosses, and other representative points (B
 and C) are also indicated in both panels.}
\label{fig:mu_sig}
\end{center}
\end{figure}

Here we interpret these results in the framework of the SSC model,
which implies that the fluences in BATSE and EGRET bands are likely
to be positively correlated through equation~(\ref{eq:EGRET
fluence}) ($F_{\rm BATSE}=F_{\rm MeV}$ and $F_{\rm EGRET}=F_{\rm
GeV}$). We further assume that the coefficient $\eta$ ($\eta_{\rm syn}$
for prompt and $\eta_{\rm IC}$ for afterglow phases) follows
some probability distribution function $p(\eta)$ which is independent
of $F_{\rm BATSE}$.  We consider a log-normal distribution with the
central value $\mu$ and standard deviation $\sigma$:
\begin{equation}
 p(\eta | \mu,\sigma) d\eta = \frac{1}{\sqrt{2\pi} \sigma} \exp
  \left[-\frac{(\log\eta - \mu)^2}{2 \sigma^2}\right]
  d\log\eta.
  \label{eq:eta distribution}
\end{equation}
Constraining  $\mu$ and $\sigma$ then leads to implications of GRB
parameters such as $\epsilon_{e}$, $\epsilon_B$, and
$\gamma_{m}$, through their relations given in the previous
section.

\begin{figure}
\begin{center}
\includegraphics[width=8.6cm]{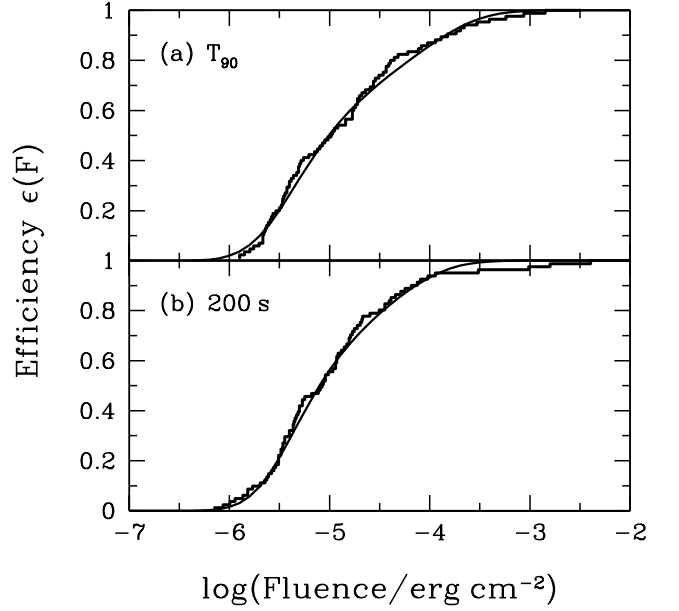}
\caption{The efficiency of EGRET for GRBs as a function of fluence,
 $\epsilon(F)$, during (a) $T_{90}$ and (b) 200 s after $T_{90}$
 (assuming a spectral index of $-2.4$). The
 histogram represents cumulative fraction of GRBs whose fluence limits
 are below a given value, which can be interpreted as the detector
 efficiency, while the solid curves are fitting function.}
 \label{fig:efficiency}
\end{center}
\end{figure}

We used the observations to constrain $\mu$ and $\sigma$ by carrying
out a maximum likelihood analysis.\footnote{The log likelihood of a
distribution is calculated by integrating the probability between
the error bars and below the upper limits of EGRET observations. For
the $T_{90}$ fluence data, we used the results of Fig.~2.3 of
\citet{Gonzalez2005} rather than Tables~2.1 and 2.2 there.}
Figure~\ref{fig:mu_sig} shows the contour plot of the most likely
region on the $\mu$--$\sigma$ plane for $T_{90}$ ({\it top}) and 200
s after $T_{90}$ data ({\it bottom}) assuming a spectral index of
$-2.4$ (if the spectral index is $-1$ then $\mu$ increases by
$\approx$1).
In that procedure, detection efficiency of EGRET as a function of
fluence, $\epsilon (F)$, is obtained from the distribution of the
EGRET upper limits (for undetected GRBs), which is shown in
Figure~\ref{fig:efficiency}; i.e., a cumulative fraction of bursts
whose fluence limits are below a given fluence.
In the case of detected GRBs, on the other hand, the size of the error
bars for the fluence is interpreted as measurement accuracy of EGRET.
Then, in order to test the consistency of the assumption that
equation (\ref{eq:eta distribution}) fits the data, we carried out a
Monte Carlo simulation that draws $10^5$ realizations of EGRET
observations assuming that the distribution of $F_{\rm EGRET} /
F_{\rm BATSE}$ follows equation (\ref{eq:eta distribution}) with the
most likely values of $\mu$ and $\sigma$.
By comparing the likelihood of these Monte Carlo realizations with
that of the actual EGRET observations, we find that 70\% of the
realizations have a lower likelihood, suggesting that equation
(\ref{eq:eta distribution}) with its most likely values is indeed
consistent with the observations.

\begin{figure}
\begin{center}
\includegraphics[width=8.6cm]{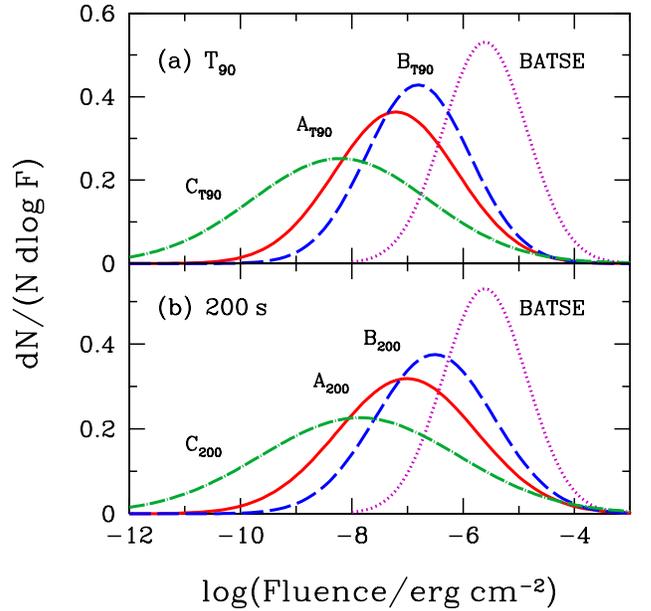}
\caption{Distribution of EGRET fluences during (a) $T_{90}$ and (b) 200
 s after $T_{90}$. Models A--C correspond to the points on
 $\mu$--$\sigma$ plots in Fig.~\ref{fig:mu_sig}. The BATSE fluence
 distribution is also plotted for comparison. The distribution for the
 prompt phase (a) is for the high-energy tail of the synchrotron
 radiation. The prompt IC fluence may be larger by up to about one order
 of magnitude (see text).}
\label{fig:dndf_egret}
\end{center}
\end{figure}

Given $\mu$ and $\sigma$, we can obtain the distribution of fluence in
EGRET band by convolving BATSE fluence distribution
($dN / dF_{\rm BATSE}$; Fig.~\ref{fig:fluence}) and
$p(\eta|\mu,\sigma)$:
\begin{equation}
 \frac{dN}{dF_{\rm EGRET}}
  = \int_{0}^{\infty}d\eta\ p(\eta|\mu,\sigma)
  \left.\frac{dN}{dF_{\rm BATSE}}\right|
  _{\eta^{-1}F_{\rm EGRET}}.
  \label{eq:fluence distribution}
\end{equation}
As representative models, we use three sets of $(\mu,\sigma)$ for both
the prompt and afterglow cases.
These are labeled as A$_{T{90}}$, B$_{T{90}}$, and C$_{T{90}}$ (A$_{\rm
200}$, B$_{\rm 200}$, and C$_{\rm 200}$), and shown in
Figure~\ref{fig:mu_sig}.
In Figure~\ref{fig:dndf_egret}, we show the resulting fluence distribution
corresponding to each of these models.

\begin{figure}
\begin{center}
\includegraphics[width=8.6cm]{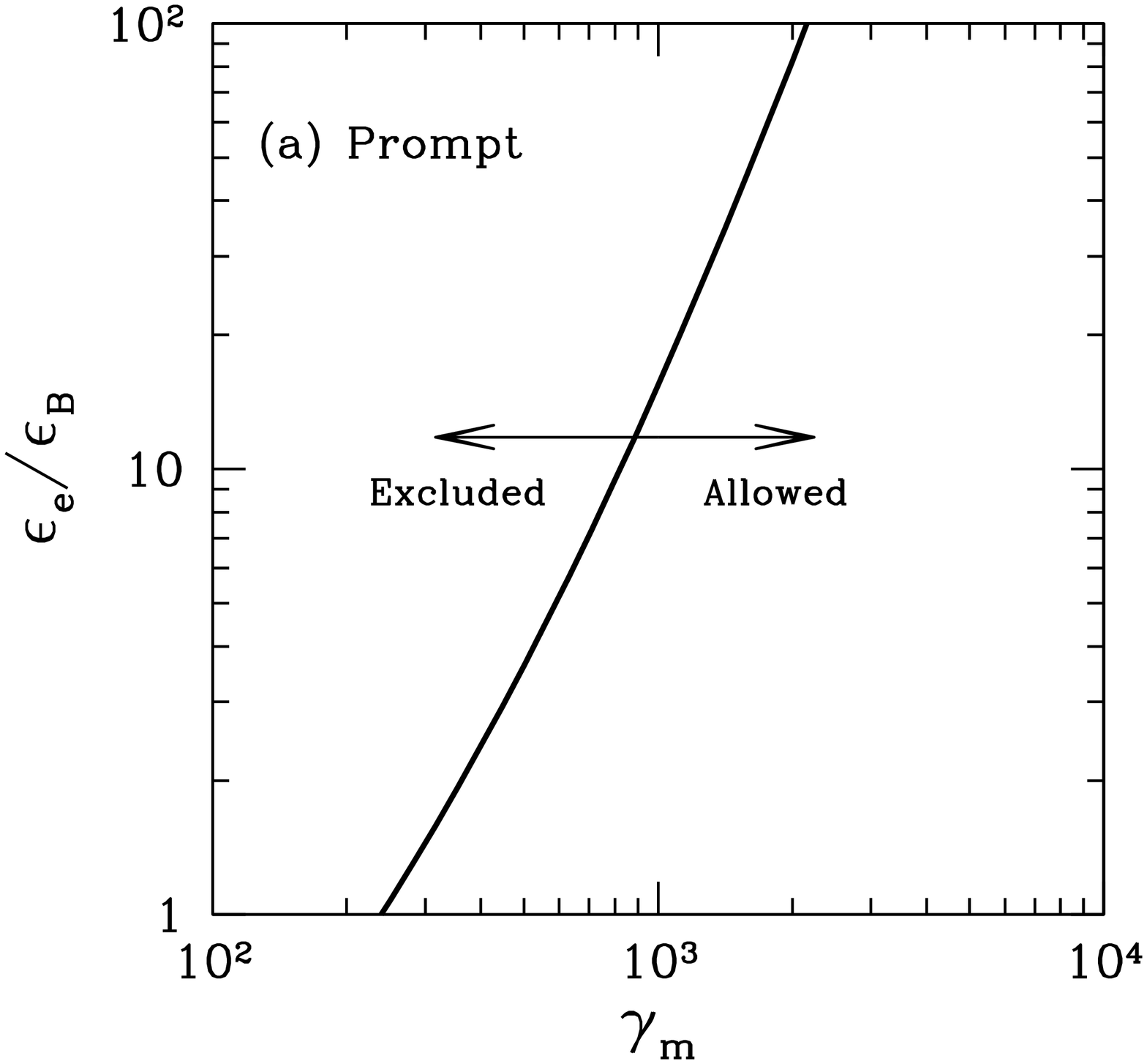}
\includegraphics[width=8.6cm]{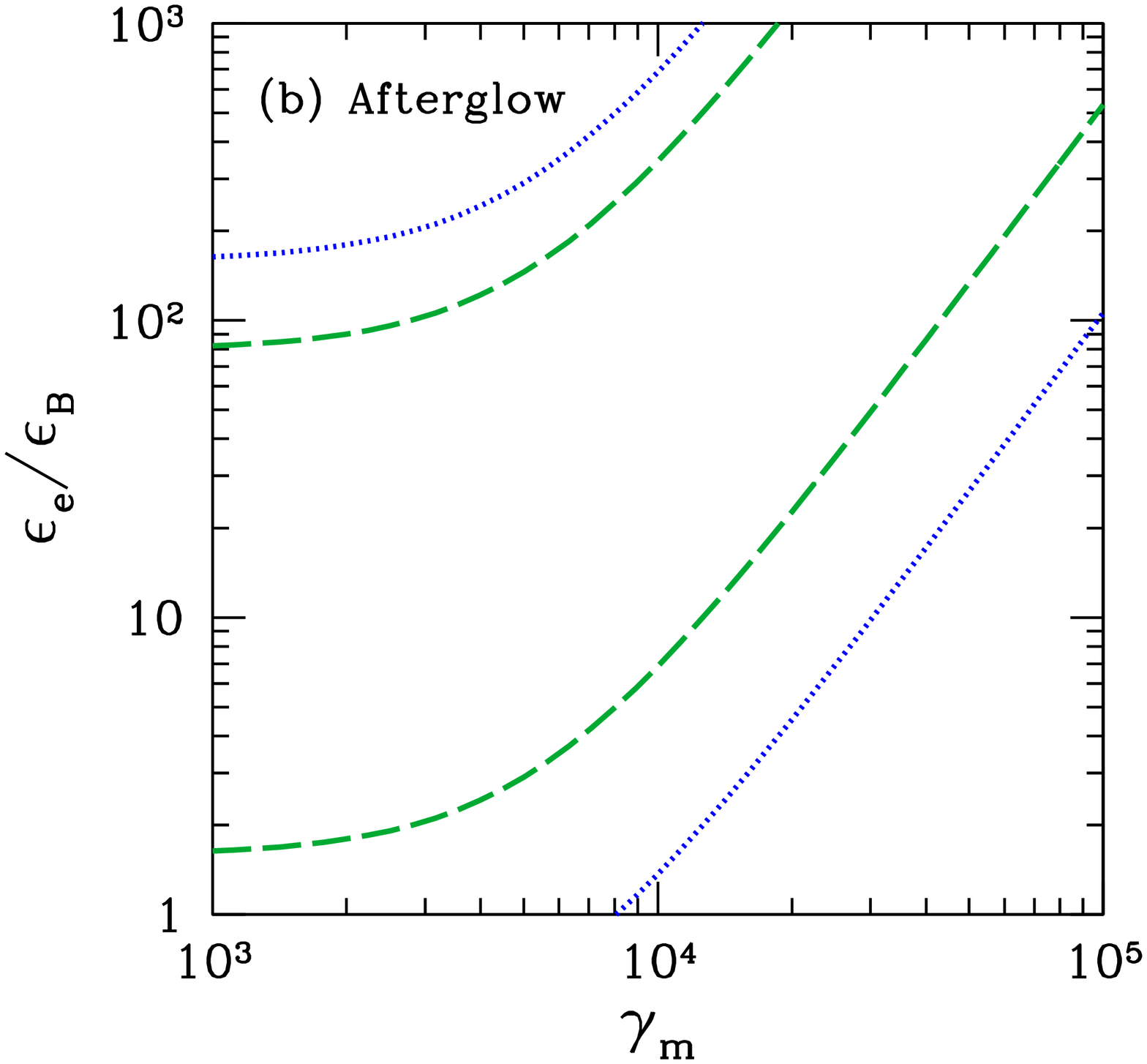}
\caption{{\it Illustrative} constraint plot on $\epsilon_e/\epsilon_B$
 and $\gamma_m$ from EGRET data for (a) prompt and (b) afterglow
 phases, obtained with canonical values for other parameters.  The left
 and right regions of the solid curve in panel (a) is excluded and
 allowed regions respectively, and regions between two dashed (dotted)
 curves in panel (b) show allowed regions corresponding to $0.013 <
 \eta_{\rm IC} < 0.09$ ($0.006 < \eta_{\rm IC} < 0.13$). Note, however,
 that these regions could easily change depending on values of other
 parameters.}
\label{fig:const}
\end{center}
\end{figure}

EGRET results imply that during the prompt emission phase,
$0.003\lesssim \eta \lesssim 0.06$. As we discussed in
\S~\ref{sub:prompt}, the low number of photons in the bursts
detected by EGRET, as well as their spectrum, implies that the
detections of prompt photons are most likely to have been dominated
by the high-energy tail of the synchrotron emission; i.e., $\eta
\approx \eta_{\rm syn}$ in Figure~\ref{fig:mu_sig}({\it top}). In
fact, simply extrapolating synchrotron tail of many BATSE bursts up
to $\sim$100 MeV regime, using inferred values for their
$\nu_{\rm syn}$ and $\alpha_2$, gives a value of $\eta_{\rm syn}$ which
is consistent with the one obtained here for the prompt phase.
The harder IC prompt emission, however, can still have as much as 10
times larger fluence than that of the synchrotron emission in EGRET
window, without being detected. Therefore, this figure also sets an
upper limit on the ratio of the IC and synchrotron components of
$\eta_{\rm IC} \lesssim 0.6$, as larger $\eta_{\rm IC}$ gives enough
photon fluence detectable by EGRET. As we showed in
\S~\ref{sub:prompt}, theoretically we predict $\eta_{\rm IC} \approx
1.2$ (for EGRET) with a canonical set of parameters.
Although this appears to imply that the current bound from EGRET
already excludes the canonical model, we cannot make such a strong
statement given the current uncertainties of many relevant parameters.
Therefore, a more conservative statement would be that the current EGRET
bound is barely consistent with the predictions of the SSC within the
internal shock model. We may interpret the bound
$\eta_{\rm IC} \lesssim 0.6$ as constraints on $\epsilon_e /
\epsilon_B$ and $\gamma_m$, which is shown in
Figure~\ref{fig:const}(a). As the Klein-Nishina suppression
($\xi_{\rm KN}$) becomes significant for large $\gamma_m$, we have
only modest limit on $\epsilon_e / \epsilon_B$ in such a regime.
However, one should keep in mind that these are order of
magnitude constraints, which may farther vary with other parameters,
such as $\nu_{\rm syn}, \alpha_2$ and $\Gamma_b$. Much better
constraint plot is expected with the future {\it GLAST} data, where
hopefully, $\eta_{\rm IC}$ will be measured for many individual bursts.

During the afterglow the synchrotron emission is much softer than
during the prompt phase, and therefore, the IC component is expected
to dominate EGRET observations also near its lower energy-band
limit. Moreover, the fact that the number of bursts detected by
EGRET during the afterglow is higher than the number detected during
the prompt emission suggests that here EGRET is likely to have
detected the actual IC component of the afterglow. The spectral
index of the GeV afterglow in EGRET window during the first 200 s is
expected to be $\alpha = 1.5$--2, implying that the evaluation of
$\mu$ in the bottom panel of Figure~\ref{fig:mu_sig}, which assumes
a spectral index of $-2.4$, might be larger by at most a small
factor ($\sim$2--3). Thus, for the afterglow, $\eta_{\rm IC} \sim
0.01$--$0.1$. We then compare this result with the theoretical
expectation of $\eta_{\rm IC}$ in equation~(\ref{eq:eta_IC}).
But first we need to estimate the value of $F_{\rm syn}/F_{\rm MeV}$
where $F_{\rm syn}$ is measured during the first 200 s following $T_{90}$
and $F_{\rm MeV}$ is the prompt emission fluence. We use the {\it Swift} GRB
table\footnote{http://swift.gsfc.nasa.gov/docs/swift/archive/grb\_table/
} which provides X-ray afterglow fluences  several tens to several
hundreds of seconds after the bursts, as well as the prompt MeV
fluences.  Using only bursts where the X-ray observation starts
after $T_{90}$ but no more than $300$ s after the burst we find a
distribution of $F_{\rm syn} / F_{\rm MeV}$ that ranges from
$10^{-3}$ to $0.1$, with the central value of $\sim$10$^{-2}$. Thus
afterglow theory with canonical parameters predicts $\eta_{\rm IC} \sim
10^{-2}$ with a large spread, consistent with EGRET constraints.
Figure~\ref{fig:const}(b)  shows the interpretation
of EGRET constraint on $\eta_{\rm IC}$ (Fig.~\ref{fig:mu_sig}) as that
for $\epsilon_e / \epsilon_B$ and $\gamma_m$, assuming canonical
parameters and $F_{\rm syn} / F_{\rm MeV} = 10^{-2}$. Although this
allowed region may change with other model parameters, again one
cannot have too large value of $\gamma_m$ because of the
Klein-Nishina suppression factor $\xi_{\rm KN}$.

\section{Implication for \textit{GLAST}}
\label{sec:GLAST}

\begin{deluxetable}{cccc}
\tablecaption{{\it GLAST}-LAT Fluence Sensitivity}
\tablewidth{8.6cm}
\tablehead{
 \colhead{$\alpha$} & \colhead{$t_0$ [s]} & \colhead{$F_{\rm lim} (t \le
 t_0)$ [erg cm$^{-2}$]} & \colhead{$F_{\rm lim}(10^3~{\rm s})$ [erg
 cm$^{-2}$]}
 }
 \startdata
 2.3 & 650  & $4.5 \times 10^{-7}$ & $5.6 \times 10^{-7}$\\
 2.0 & 650  & $6.6 \times 10^{-7}$ & $8.1 \times 10^{-7}$\\
 1.0\tablenotemark{a} & 650  & $5.2 \times 10^{-6}$ & $6.4 \times
 10^{-6}$\\
\enddata
\tablecomments{Parameters of point-source fluence sensitivity (integrated
 over 30 MeV--30 GeV) of {\it GLAST}-LAT (see eq.~[\ref{eq:GLAST
 limit}]). The power-law index is $-\alpha$, and the unit in fluence
 limit $F_{\rm lim}$ is erg cm$^{-2}$. The detection criterion for $t
 \le t_0$ is five photons, and significance for $t > t_0$ is $5\sigma$,
 where $t_0 = 650$ s is the transition time.}
\label{table:GLAST}
\tablenotetext{a}{Here we considered a detection based on the number of
photons in the energy range 30 MeV--30 GeV. A higher $t_0$ and more
sensitive background limited threshold can be obtained for $\alpha =
1$ if a higher energy range is considered (see text and
 Appendix~\ref{app:GLAST}).}
\end{deluxetable}

We now move on to discussions on implications for {\it GLAST} using the
obtained constraints on $\eta$ in the previous section.
First we estimate the sensitivity of LAT on board {\it GLAST} for
prompt and afterglow GeV emission, based on its published
sensitivity to steady point
sources,\footnote{http://www-glast.slac.stanford.edu/} which is $4
\times 10^{-9}$ cm$^{-2}$ s$^{-1}$ above 100 MeV at $5\sigma$ with a
power-law index of $-2$. This sensitivity is obtained by a one-year
all-sky survey during which any point source is observed for
$\sim$70 d (the LAT field of view is 2.4 sr).\footnote{We assume
here a step function for the LAT window function.} Therefore during
the background-limited regime (when $t$ is large enough that many
background photons are observed) the flux limit scale with $t$ as $4
\times 10^{-9}$ cm$^{-2}$ s$^{-1}$ $(t/70~\mathrm{d})^{-1/2}$.
During the photon-count-limited regime (when $t$ is so small that less
than one background photon is expected), in contrast, the detection
limit is at a constant fluence. Therefore the fluence sensitivity of the
{\it GLAST}-LAT detector is
\begin{equation}
 F_{\rm lim}(t) \approx
  \left\{
   \begin{array}{ll}
    F_{\rm lim} (t_0) & [t \le t_0],\\
    F_{\rm lim} (t_0) \left(\frac{t}{t_0}\right)^{1/2} & [t > t_0],
   \end{array}
  \right.
  \label{eq:GLAST limit}
\end{equation}
where $t_0 = 650$ s represents the time when the transition from
photon-count-limited to background-limited regime occurs in the LAT
case. Note that equation~(\ref{eq:GLAST limit}) is for the limiting {\it
fluence}, the time-integrated flux, rather than the flux. This limit
is more natural in the photon-count-limited regime and it is more
relevant to EGRET constraints that we derived in the previous section.
Detailed derivation of this sensitivity is given in
Appendix~\ref{app:GLAST}. In Table~\ref{table:GLAST}, we summarize
the values of $t_0$ and $F_{\rm lim}(t)$ for a few cases of power
law index $-\alpha$ and integration time $t$. The values of $F_{\rm
lim}(t)$  for $t \ll t_0$ in the table are determined by criteria of
five-photon detection, while those for $t > t_0$ are by $5\sigma$
significance. The fluence we argue here is the one integrated over 30
MeV--30 GeV, in order to compare with the EGRET fluence upper bounds.

\begin{deluxetable}{ccc}
\tablewidth{8.6cm}
\tablecaption{GRB Rate at {\it GLAST}-LAT and Contribution to the EGB
 Flux}
\tablehead{
 \colhead{Model} & \colhead{Rate at {\it GLAST}} &  \colhead{$I_{\rm
 EGB}$ [GeV cm$^{-2}$ s$^{-1}$ sr$^{-1}$]}}
\startdata
 A$_{T{90}}$ & 15 yr$^{-1}$ & $6.3 \times 10^{-10} ~(1+\eta_{\rm IC} /
 \eta_{\rm syn}$) \\
 B$_{T{90}}$ & 20 yr$^{-1}$ & $8.4 \times 10^{-10} ~(1+\eta_{\rm IC} /
 \eta_{\rm syn}$) \\
 C$_{T{90}}$ & 10 yr$^{-1}$ & $4.4 \times 10^{-10} ~(1+\eta_{\rm IC} /
 \eta_{\rm syn}$) \\
 \hline
 A$_{200}$ & 20 yr$^{-1}$ & $8.9 \times 10^{-10}$ \\
 B$_{200}$ & 30 yr$^{-1}$ & $1.3 \times 10^{-9}$ \\
 C$_{200}$ & 15 yr$^{-1}$ & $6.5 \times 10^{-10}$ \\
\enddata
\tablecomments{The estimate of detection rate with {\it GLAST}-LAT
 (for $\alpha = 2.3$), and expected EGB intensity, for models A, B, and
 C of the prompt (during $T_{90}$) and afterglow phases (during 200 s
 after $T_{90}$).  The correction factor $1+\eta_{\rm IC} / \eta_{\rm
 syn}$ for $I_{\rm EGB}$ in the case of prompt emission could be as
 large as $\sim$10.  Also note that these estimates are quite
 conservative.  See discussions in \S\S~\ref{sec:GLAST} and
 \ref{sec:EGB} for more details.}
\label{table:result}
\end{deluxetable}

In the case of background-limited regime, it might be more
appropriate to use higher energy threshold (instead of 30 MeV)
especially for hard source spectrum, because the background spectrum
falls steeply with frequency ($\alpha \simeq 2.1$). We may find
optimal low-frequency threshold depending on spectral index of GRB
emissions; it is higher for harder spectrum. Thus, we should be able
to improve the fluence sensitivity for background-limited regime,
compared with the figures given in Table~\ref{table:GLAST}. In
addition, transition from photon-count to background limited regime
would occur later than 650 s. For our purpose, however, as time
scales we consider ($T_{90}$ for prompt emission and 200 s after
$T_{90}$ for afterglows) are both during photon-count-limited
regime, the consideration above does not apply and we can use full
energy range (30 MeV--30 GeV for EGRET) to collect as many photons as
possible.

{\it GLAST} is also equipped with the GLAST Burst Monitor (GBM)
instrument, dedicated for the detection of GRBs. It detects photons
of 8 keV to more than 25 MeV and its field of view is $\sim$8 sr.
The expected rate of GRBs that trigger GBM is $\sim$200 yr$^{-1}$
\citep{McEnery2006}, which is almost as high as BATSE rate. Each
year, about 70 out of these $\sim$200 bursts should fall within the
LAT field of view. Given the distribution of fluences
(Fig.~\ref{fig:dndf_egret}) and the LAT sensitivity
(Table~\ref{table:GLAST}), we can estimate the fraction of GRBs that
would be detected with LAT. In Table~\ref{table:result}, we show the
expected LAT detection rate for $\alpha = 2.3$, which is $\sim$20
yr$^{-1}$ for the best-fit models of the EGRET data for both the
prompt and afterglow emissions. The prompt phase estimates are for
detections of the synchrotron component in the $\sim$100 MeV range.
Given the large effective area of the LAT it is expected also to
detect $\gtrsim$ GeV photons from the IC component and identify the
spectral break associated with the transition from the synchrotron
to IC component, thereby directly testing the SSC model.

The estimates given in Table~\ref{table:result} are fairly
conservative. First, while we used five-photon criterion for the
detection, even two-photon detection should be quite significant,
because the expected background count is much smaller than one
photon during $T_{90}$ and the following 200 s that we considered.
Second, {\it Swift} can find dimmer bursts than GBM. Although the
discovery rate is not as high as that of GBM or BATSE, it would
still be able to find tens of new GRBs in the LAT field of view.
Thus the true rate would likely be larger than the figures given in
Table~\ref{table:result}.

\section{Implication for the extragalactic gamma-ray
background}
\label{sec:EGB}

\begin{figure}
\begin{center}
\includegraphics[width=8.6cm]{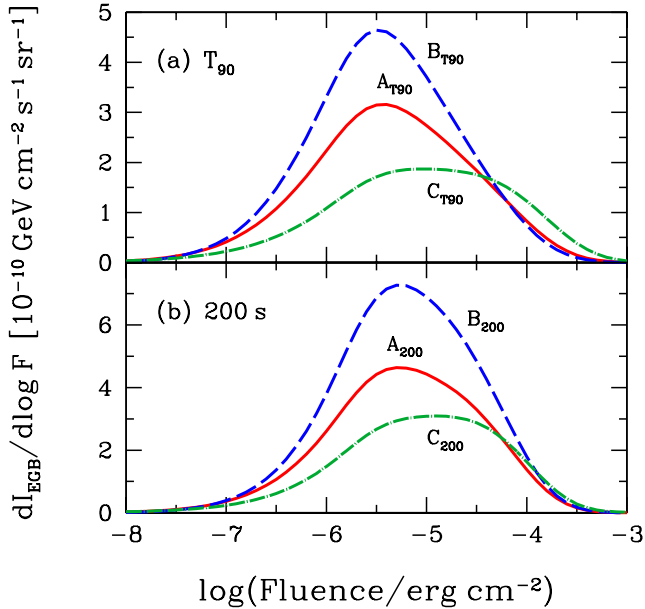}
\caption{Contribution to the EGB intensity $I_{\rm EGB}$ from GRBs of
 given fluence, for (a) prompt (during $T_{90}$) and (b) afterglow
 emission (200 s after $T_{90}$) phases. In each panel, three models
 A--C are shown. Note that for the prompt phase, the fluence is that for
 synchrotron radiation, and that for IC component could be even larger
 (see text).}
\label{fig:egb_dist}
\end{center}
\end{figure}

All the GRBs except for those detected by EGRET should contribute to
the EGB flux to a certain extent \citep{Dermer2007}. This may be
computed as
\begin{equation}
 I_{\rm EGB}  =  \frac{R_{\rm GRB}}{4 \pi}
  \int_{0}^{\infty} dF\ F\frac{dP}{dF}
  \left[ 1 - \epsilon(F) \right],
  \label{eq:EGB}
\end{equation}
where $F$ is EGRET fluence in 30 MeV--30 GeV, $dP / dF$ is the
normalized distribution of EGRET fluence
(eq.~[\ref{eq:fluence distribution}] and
Fig.~\ref{fig:dndf_egret}), and $R_{\rm GRB} \sim 2$~d$^{-1}$ is the
occurrence rate of GRBs from all sky. The factor $1 - \epsilon (F)$
takes into account the fact that very bright GRBs cannot contribute
to the EGB because they would be identified as point sources (but
see discussions below). Figure~\ref{fig:egb_dist} shows differential
EGB intensity $dI_{\rm EGB} / d\log F$ that
represents contribution from GRBs of a given fluence, for prompt and
afterglow phases. In the third column of Table~\ref{table:result},
we show the EGB intensity due to prompt and afterglow phases of GRBs
which is $\sim$10$^{-9}$ GeV cm$^{-2}$ s$^{-1}$ sr$^{-1}$. On the
other hand, in the same energy range, EGRET measured the EGB flux to
be $10^{-5}$ GeV cm$^{-2}$ s$^{-1}$ sr$^{-1}$ \citep{Sreekumar1998}.
Therefore, GRBs that were detected by BATSE but were not detected as
point sources by EGRET contribute to the EGB at least $\sim$0.01\%.
Again, we note that the estimates for the prompt phase are those of
synchrotron component. We thus need to take the predicted IC
contribution into account, which is represented by a correction factor
$1+\eta_{\rm IC} / \eta_{\rm syn}$ in Table~\ref{table:result}.
Since this factor could be as large as $\sim$10 according to the
discussion in \S~\ref{sec:Constraint on high-energy emission with
EGRET}, EGB flux due to prompt phase of GRBs could also becomes
$\sim$10 times larger, which makes GRB contribution as large as
$\sim$0.1\% of the observations above $\sim$GeV. In any case, the
contributions from other astrophysical sources such as blazars are
expected to be more significant than GRBs \cite[e.g,][and references
therein]{Ando2007b}.

Additional contribution to EGB is expected from a large number of
GRBs that point away from us and therefore would not have been
detected with BATSE. The emission from these bursts points towards
us once the external shock decelerates \citep{Rhoads1997}. Since the
total GeV energy emitted every decade of time during the afterglow
is roughly constant, the contribution of these GRBs to EGB can be
estimated by the GeV emission of the bursts that were detected by
BATSE. Similar contribution is expected from bursts that points
towards us but that are too faint to be detected by BATSE, if the
GRB luminosity function behaves as $\phi(L) \propto L^{-2}$ as
suggested by the universal structured jet model \citep*[][see
however \citealt*{Guetta05}]{Lipunov01,Rossi02,Zhang02,Perna2003}.
Therefore the contribution of bursts that were not detected by BATSE
to EGB can be estimated by the afterglow fluence of the detected
bursts, assuming no contribution from bursts with only an upper limit.
This is a reasonable estimate since the GeV flux is dominated by the
few brightest bursts in GeV which are the most likely to be
detected. Taking the fluence of the detected GeV bursts as the
logarithmic mean of these upper and lower limits implies $I_{\rm
EGB} \sim 5 \times 10^{-9}$ GeV cm$^{-2}$ s$^{-1}$ sr$^{-1}$, a GRB
contribution being $\sim$0.1\% of the EGB.

Finally, we note that there is a big uncertainty in removing the
Galactic foreground contamination from the total diffuse flux
\citep{Keshet2004b}. Additionally, EGRET observations do not
constrain TeV emission that cascades down into the GeV range for
GRBs at cosmological distances \citep{Casanova07, Murase2007}. Thus,
if the foreground subtraction was indeed underestimated or if GRB TeV
emission is not negligible, then GRB contribution might be much more
significant than the estimates here.

\section{Summary and Conclusions}
\label{sec:conclusions}

The {\it GLAST} satellite would enable us to test
high-energy emission mechanisms of GRBs. If this emission will be
found to be consistent with SSC then its observations would
constrain physical parameters such as $\epsilon_{e}/\epsilon_B$
ratio and the bulk Lorentz factor of the jet, $\Gamma_{b}$. The
EGRET instrument on board {\it CGRO}, while less sensitive than the
{\it GLAST}-LAT detector, identified several BATSE GRBs with GeV
photons. In addition, stringent upper limits for $\sim$100 GRBs were
put on fluences in the GeV band by analyzing the EGRET data
\citep{Gonzalez2005}.

In this paper, we further extended this EGRET result, comparing with
the SSC emission model. Following theoretical models of SSC, we assumed
that there is a linear correlation between fluences in
BATSE and EGRET energy bands, and that the proportionality
coefficient $\eta$ follows a log-normal distribution. We found that
the predictions from the SSC model using canonical parameter values is
fully consistent with EGRET fluence measurements and upper limits for
both the prompt and afterglow phases. During the course of showing
this result, we properly took the Klein-Nishina feedback effect into
account in the theoretical calculation. The
best-fit value of the coefficient was $\log \eta \simeq -1.5$ for
both the prompt and afterglow emissions, and it is already stringent
enough to test the SSC scenario.
The limits for the prompt emission phase are for the synchrotron
radiation, and thus if we consider the IC component as well, the value
of $\eta$ could be larger by up to one order of magnitude.

The obtained $\eta$ distribution, together with the BATSE fluence
distribution, gives the expected fluence distribution in the GeV
band, which is shown in Figure~\ref{fig:dndf_egret}. As the {\it
GLAST}-LAT detector covers EGRET energy band, we can predict the
detectable number of GRBs with {\it GLAST} from the distribution of
$F_{\rm EGRET}$, given the {\it GLAST}-LAT sensitivity. Our
conservative estimate using the five-photon criterion is that about
$\sim$20 GRBs among those detected with GBM would be detected with
{\it GLAST}-LAT each year. This number could be even larger if we
use fewer-photon criteria.
The fluence distribution can also be used to estimate the GRB
contribution to the EGB intensity. We found that the contribution
would be at least $\sim$0.01\% but is likely to be as large as
$\sim$0.1\%.

\acknowledgments

We are grateful to B.~L. Dingus and M.~M. Gonz{\'a}lez S{\'a}nchez
for very helpful comments and discussions.  We thank the referee for
useful comments. This work was supported by Sherman Fairchild
Foundation (SA and EN), NASA Swift Grant (EN), Alfred P. Sloan
Foundation, Packard Foundation, and a NASA ATP Grant (RS).

\appendix

\section{Klein-Nishina feedback on high-energy emission}
\label{app:KN}

We shall find an analytic expression for $\xi_{\rm KN}$ due to the
Klein-Nishina feedback.
To simplify the argument such that we can treat it analytically, we make
the following approximations: (i) an electron with a fixed Lorentz
factor $\gamma_e$ radiates mono-energetic synchrotron photons; (ii) the
same electron upscatter a given synchrotron photon to another
monochromatic energy, which is increased by a factor of $\gamma_e^2$;
(iii) $\nu f_\nu$ of both synchrotron and IC photons peaks at
$\nu_{\rm syn}$ (a synchrotron frequency corresponding to $\gamma_m$)
and $\nu_{\rm IC} (= \gamma_m^2 \nu_{\rm syn}$; if there is no
Klein-Nishina suppression), respectively; (iv)
the Klein-Nishina cutoff occurs quite sharply above its threshold; (v)
both cooling and self-absorption frequencies are much smaller than the
frequency region of our interest; and (vi) electrons cool so quickly
that any dynamical effects can be neglected.
With these approximations, expressions for the ratio of power of
synchrotron and IC radiations from a given electron $Y(\gamma_e) =
P_{\rm IC}(\gamma_e) / P_{\rm syn}(\gamma_e)$ simplifies significantly.
In particular, according to the assumption (iii) above, we have
$(\epsilon_e / \epsilon_B)^{1/2} \xi_{\rm KN} \approx Y(\gamma_m)$.
This is given as
\begin{equation}
 Y({\gamma_m}) = \frac{\epsilon_e}{\epsilon_B}
  \frac{p/2-1}{p-1} (\nu_{\rm syn}^{\prime})^{p/2-1}
  \int_0^\infty d\nu^\prime
  \frac{\max[\nu^\prime,\nu_{\rm syn}^\prime]^{-(p-1)/2}
  (\nu^\prime)^{-1/2}}
  {1+Y([\nu^\prime / \nu_{\rm syn}^\prime]^{1/2}\gamma_m)}
  \Theta\left(-\nu^\prime + \frac{m_ec^2}{h\gamma_m}\right),
  \label{eq:Y}
\end{equation}
where $p$ is electron spectral index, $\Theta$ is the step function, and
primed quantities are evaluated in the rest frame of the ejecta (e.g.,
$\nu^\prime = \nu / \Gamma_b$, where $\nu$ is the frequency in an
observer frame).

A detailed derivation as well as numerical approaches are given
elsewhere (Nakar, Ando, \& Sari, in preparation), but at least this
equation can be understood qualitatively. For a given electron with
Lorentz factor $\gamma_m$, the synchrotron power does not depend on
whether the Klein-Nishina suppression is effective or not.
On the other hand, the IC power does, because it is proportional to
the energy density of seed (synchrotron) photons integrated up to some
cutoff frequency; synchrotron photons above this frequency cannot be
IC scattered efficiently by the electron with $\gamma_m$ because of
the Klein-Nishina suppression.
The integrand of equation~(\ref{eq:Y}) represents the synchrotron
spectrum.
More specifically, assuming there is no Klein-Nishina suppression, the
spectrum is simply given by $f_{\nu^\prime} \propto \max[\nu^\prime,
\nu_{\rm syn}^\prime]^{-(p-1)/2} (\nu^\prime)^{-1/2}$; the step function
then represents the
Klein-Nishina cutoff. The factor $1+Y$ in the denominator of the
integrand accounts for the suppression of the electron distribution
function due to the enhanced IC cooling; i.e., $dN_e / d\gamma_e \propto
(d\gamma_e / dt)^{-1} \propto [P_{\rm syn}(\gamma_e) + P_{\rm
IC}(\gamma_e)]^{-1} \propto [1 + Y(\gamma_e)]^{-1}$.
These electrons are ones that emit synchrotron photons of a given
frequency $\nu^\prime$.
Recalling the relation $\gamma_e \propto \nu^{\prime 1/2}$, their
Lorentz factor is given by $(\nu^\prime / \nu_{\rm syn}^\prime)^{1/2}
\gamma_m$, which appears in the argument of $Y$ in the integrand.
Finally, the other constants in equation~(\ref{eq:Y}) are chosen so
that we have a proper relation for the fast cooling, $Y(1+Y) =
\epsilon_e / \epsilon_B$, if we turn off the Klein-Nishina cutoff and
have constant $Y$.

\begin{figure}
\begin{center}
\includegraphics[width=8.6cm]{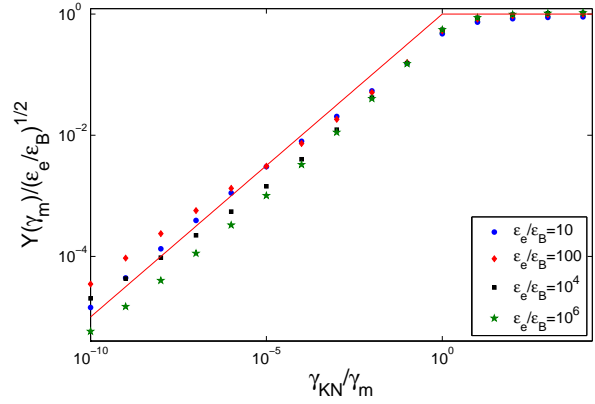}
\caption{Ratio of IC to synchrotron power $Y$ by an electron with
 Lorentz factor $\gamma_m$ as a function of $\gamma_{\rm KN} /
 \gamma_m$. Points represent numerical solutions of eq.~(\ref{eq:Y})
 for various values of $\epsilon_e/\epsilon_B$, and solid line is an
 analytic fit (eq.~[\ref{eq:eta KN}]).}
\label{fig:Ym}
\end{center}
\end{figure}

Now we shall find analytic expressions of equation~(\ref{eq:Y}) in
asymptotic regions.
We start from the case of $\gamma_m \lesssim \gamma_{\rm KN} = m_ec^2 /
h \nu_{\rm syn}^\prime$, which is equivalent to $\nu_{\rm syn}^\prime
< m_ec^2 / h \gamma_m$.
The integration then becomes
\begin{equation}
 Y(\gamma_m) = \frac{\epsilon_e}{\epsilon_B}\frac{p/2-1}{p-1}
  \left[
   \int_0^{\nu_{\rm syn}^\prime} d\nu^\prime
   \frac{(\nu_{\rm syn}^\prime \nu^\prime)^{-1/2}}
   {1+Y([\nu^\prime/\nu_{\rm syn}^\prime]^{1/2} \gamma_m)}
   +\int_{\nu_{\rm syn}^\prime}^{\frac{m_ec^2}{h\gamma_m}}
   d\nu^\prime \frac{(\nu_{\rm syn}^\prime)^{p/2-1}
  (\nu^\prime)^{-p/2}}{1+Y([\nu^\prime/\nu_{\rm
  syn}^\prime]^{1/2} \gamma_m)}
  \right].
\end{equation}
We assume that the function $1 + Y$ varies rather mildly in the
integrand, so that in the argument of $Y$ we may use $\nu^\prime =
\nu_{\rm syn}^\prime$.
Then the integral can be evaluated analytically, and gives $Y(\gamma_m)
[1 + Y(\gamma_m)] = \epsilon_e / \epsilon_B$.
When $\epsilon_e \gg \epsilon_B$, we have $Y(\gamma_m) = (\epsilon_e /
\epsilon_B)^{1/2}$, which is the same result as in the case of no
Klein-Nishina suppression.
This makes sense because the condition $\gamma_m < \gamma_{\rm KN}$
indicates that the electrons with $\gamma_m$ is below the Klein-Nishina
threshold with seed photons at frequency $\nu_{\rm syn}^\prime$ that
dominate the synchrotron power.
On the other hand, when $\gamma_m > \gamma_{\rm KN}$ (or $\nu_{\rm
syn}^\prime > m_ec^2/h\gamma_m$), equation~(\ref{eq:Y}) becomes
\begin{equation}
 Y(\gamma_m) = \frac{\epsilon_e}{\epsilon_B}\frac{p/2-1}{p-1}
   \int_{0}^{\frac{m_ec^2}{h\gamma_m}}
   d\nu^\prime \frac{(\nu_{\rm syn}^\prime \nu^\prime)^{-1/2}}
   {1+Y([\nu^\prime/\nu_{\rm syn}^\prime]^{1/2}\gamma_m)}
   \approx \frac{\epsilon_e}{\epsilon_B}\frac{p-2}{p-1}
   \left(\frac{\gamma_m}{\gamma_{\rm KN}}\right)^{-1/2}
   \frac{1}{1+Y([\gamma_m\gamma_{\rm KN}]^{1/2})},
   \label{eq:Y for large gamma_m}
\end{equation}
where in the second equality, we used $\nu^\prime = m_ec^2/h\gamma_m$
for the argument of $Y$.
When $\gamma_m / \gamma_{\rm KN}$ is large enough so that $Y([\gamma_m
  \gamma_{\rm KN}]^{1/2}) \ll
1$, then equation~(\ref{eq:Y for large gamma_m}) immediately gives
asymptotic solution for $Y(\gamma_m)$.
When $\gamma_m$ is in the intermediate regime, we can still get analytic
expressions, which however are given elsewhere because they are somewhat
complicated.
Here we simply show numerical solutions of equation~(\ref{eq:Y}) as a
function of $\gamma_{\rm KN} / \gamma_m$ for various values of
$\epsilon_e / \epsilon_B$.
We show these results as well as a simple fitting form (given by
eq.~[\ref{eq:eta KN}]) in Figure~\ref{fig:Ym}.
Thus, equation~(\ref{eq:eta KN}) provides fairly good fit to the results
of numerical integration of equation~(\ref{eq:Y}).

\section{Fluence sensitivity of \textit{GLAST}}
\label{app:GLAST}

For a steady point source with a spectral index of $-2$, the
sensitivity of {\it GLAST}-LAT to its flux above 100 MeV is $4
\times 10^{-9}$ cm$^{-2}$ s$^{-1}$ at $5\sigma$ significance for a
one-year all-sky survey. Considering the field of view of {\it
GLAST}-LAT, 2.4 sr, this survey time corresponds to 70 d exposure
time to the source, and therefore, the sensitivity to the number
fluence integrated over this time scale is $2.4 \times 10^{-2}$
cm$^{-2}$. In this section, we generalize this limit to an arbitrary
spectral index, $-\alpha$, and exposure time, $t$.

Before starting the discussion, we define the differential number
and energy fluences, and integrated number and energy fluences (all
quantities are time-integrated):
\begin{equation}
 \frac{dF_{N}}{dE}  =  C E^{-\alpha},
 ~~
 \frac{dF}{dE}  =  C E^{1-\alpha},
  \label{eq:dNdE}
\end{equation}
\begin{equation}
 F_{N}  =  C \frac{E_{\rm max}^{1-\alpha} - E_{\rm
  min}^{1-\alpha}} {1-\alpha},
  ~~
 F  =  C \frac{E_{\rm max}^{2-\alpha} - E_{\rm
  min}^{2-\alpha}} {2-\alpha},
  \label{eq:N}
\end{equation}
where $C$ is a coefficient, and $E_{\rm min}$ and $E_{\rm max}$ are
the energy band boundaries.

The fluence sensitivity for a one year exposure is within the
background-limited regime---namely within one year many
background photons are expected to be detected within the
point-spread-function of the detector. In the case of {\it
GLAST}-LAT, backgrounds are the EGB or Galactic foreground
emissions. Therefore, we start our discussion from this
background-limited case. Let us define this background rate of {\it
GLAST} by $\dot N_{\rm bg}$, for which we assume $E^{-2.1}$ spectrum
and use the energy-dependent angular resolution and on-source
effective area $A_{\rm eff}(E)$.\footnote{For these specifications
of the detector, we use the results shown in
http://www-glast.stanford.edu/} The criterion of point-source
detection is
\begin{equation}
 N_\gamma > N_{\gamma,{\rm lim}} \equiv \sigma \sqrt{\dot N_{\rm bg} t},
  \label{eq:detection criterion}
\end{equation}
where $\sigma$ represents significance of detection, and photon count
from the source is obtained by
\begin{equation}
 N_\gamma = \int_{E_{\rm min}}^{E_{\rm max}}
  dE \frac{dF_{N}}{dE} A_{\rm eff}(E).
  \label{eq:photon count}
\end{equation}
Therefore, using equation~(\ref{eq:dNdE}) in equations~(\ref{eq:photon
count}) and (\ref{eq:detection criterion}), we can obtain the
sensitivity to the coefficient $C_{\rm lim}$ as follows:
\begin{equation}
 C_{\rm lim} = N_{\gamma,{\rm lim}}
  \left[\int_{E_{\rm min}}^{E_{\rm max}} dE E^{-\alpha} A_{\rm
  eff}(E) \right]^{-1},
  \label{eq:C_lim}
\end{equation}
and then using equation~(\ref{eq:N}), this can be
translated into the sensitivity to the number and energy fluences,
$F_{N,{\rm lim}}$ and $F_{\rm lim}$. We here note that
$C_{\rm lim}$ depends on $t$, $\alpha$, $E_{\rm min}$, and $E_{\rm
max}$, while $N_{\gamma,{\rm lim}}$ depends only on $t$, $E_{\rm
min}$, and $E_{\rm max}$. In this background-limited regime, the
time dependence is $F_{\rm lim} \propto t^{1/2}$ from
equation~(\ref{eq:detection criterion}). We confirmed that, using
the EGB intensity measured by EGRET \cite{Sreekumar1998} and
energy-dependent angular resolution of LAT, we could obtain the
limit comparable to $F_{N,{\rm lim}} = 2.4 \times 10^{-2}$
cm$^{-2}$, in the case of $\alpha = 2$, $t = 70$ d, $E_{\rm min} =
100$ MeV, $E_{\rm max} = \infty$, and $\sigma = 5$. The results of
this procedure for several values of interest of $\alpha$ are
summarized in equation~(\ref{eq:GLAST limit}) and
Table~\ref{table:GLAST}. Here, we used EGRET energy range, i.e.,
$E_{\rm min} = 30$ MeV and $E_{\rm max} = 30$ GeV, but we can instead
adopt different values.

If the time scale is short such that $N_{\gamma,{\rm lim}} < 1$,
then the argument above does not apply, but the sensitivity is
simply obtained by the expected photon count from the source. In
this photon-count-limited regime, we can evaluate the fluence
sensitivity by requiring $N_\gamma$ to be a few; here we use
$N_\gamma = 5$. One can obtain the corresponding $C_{\rm lim}$ by
solving this criterion using equation~(\ref{eq:photon count}). This
time, $C_{\rm lim}$ is independent of $t$. Then again using
equation~(\ref{eq:N}), one can get the fluence
sensitivity in this regime as shown in Table~\ref{table:GLAST}.

\end{document}